\documentclass[a4paper,11pt]{article}

\usepackage{jcappub}

\usepackage{slashed}
\usepackage{youngtab}

\newcommand\fverb{\setbox\fverbbox=\hbox\bgroup\verb}
\newcommand\fverbdo{\egroup\medskip\noindent%
            \fbox{\unhbox\fverbbox}\ }
\newcommand\fverbit{\egroup\item[\fbox{\unhbox\fverbbox}]}
\newbox\fverbbox
\newcommand{\nablaslash}{\not{\hbox{\kern-3pt $\nabla$}}}

\title{\boldmath Primordial massive gravitational waves from Einstein-Chern-Simons-Weyl gravity}

\author{Yun~Soo~Myung}
\author{and~Taeyoon~Moon}
\affiliation{Institute of Basic Sciences and Department  of Computer Simulation, Inje University,\\
Gimhae 621-749, Korea}

\emailAdd{ysmyung@inje.ac.kr} \emailAdd{tymoon@inje.ac.kr}

\abstract{We investigate  the evolution of cosmological
perturbations during de Sitter inflation in the
Einstein-Chern-Simons-Weyl gravity. Primordial massive gravitational
waves are composed of one scalar, two vector and four tensor
circularly polarized  modes. We show that  the vector power spectrum
decays quickly like a transversely massive vector  in the
superhorizon limit $z\to 0$. In this limit, the power spectrum
coming from massive tensor modes decays quickly, leading to the
conventional tensor power spectrum.
 Also, we find that in the
limit of $m^2 \to 0$ (keeping the Weyl-squared term only), the
vector and tensor power spectra disappear.  It implies  that their
power spectra are not gravitationally produced because they (vector
and tensor) are decoupled from the expanding de Sitter background,
as a result of conformal invariance.}

\begin{document}

\maketitle \flushbottom

\section{Introduction}
The recent detection of  primordial gravitational waves (GWs) via
B-mode polarization of Cosmic Microwave Background Radiation (CMBR)
by BICEP2~\cite{Ade:2014xna} has shown that the cosmic  inflation at
a high scale of $10^{16}$ GeV is the most plausible source of
generating primordial GWs. The primordial GWs can be imprinted  in
the anisotropies and polarization spectrum of CMBR
 by making  the photon redshifts. The B-mode signal
observed by BICEP2 might  also contain contributions from other
sources (vector modes, cosmic strings) in addition to tensor
modes~\cite{Moss:2014cra}.

 The prediction about B-modes from inflation implies the phenomenon of GWs as well as quantum gravity.
 In order to explore this situation
explicitly, we would like to mention that no conventional experiment
is capable of detecting individual gravitons (quanta of the
gravitational field) like photons  (quanta of the electromagnetic
field) because the LIGO is supposed to detect GWs ($h_+,h_\times$)
with a strain amplitude of $10^{-21}$ which amounts to
$3\times10^{37}$ gravitons~\cite{DYSON:2013jra}. This implies that
if the LIGO wants to detect a single graviton, its sensitivity
should be improved by a factor of the order of $3\times10^{37}$. The
inflation implies  a brief period during which the universe
underwent an exponential expansion. If inflation occurred, however,
the universe affords an access to detect gravitons because the
inflation is considered as an ideal graviton amplifier to produce
primordial GWs~\cite{Krauss:2014sua}. In this way, the inflation
produces a classical signal of macroscopic GWs in response to
spontaneous emission of gravitons. A classical signal of GWs may be
considered as a coherent superposition of a large number of
gravitons. This is similar to the LASER (light amplification by
stimulated emission of radiation) which is a device that emits light
through a process of optical amplification based on the stimulated
emission of electromagnetic radiation. In this sense, one may regard
the inflation as a process of GWASG (gravitational wave
amplification by spontaneous emission of gravitons).    The
difference is that the light (GW) is amplified by stimulated
(spontaneous) emission of photons (gravitons). The simplest effect
of primordial GWs is to produce a direct quadrupole anisotropy in
the CMBR, inducing B-mode polarizations through Thomson scattering.
Furthermore, the mechanism of cosmic inflation naturally generates a
stochastic background of primordial GWs which is an incoherent
superposition of GWs~\cite{Fidler:2014oda}.

The quadrupole anisotropy usually  arises from 3 types of
cosmological perturbations in Einstein gravity: scalar (due to
density fluctuations); vector (due to vorticity induced by
defects/strings); tensor (due to gravity waves). The curl-free
E-mode  may be due to both the scalar and tensor perturbations,
whereas the B-mode is due to only vector or tensor perturbations
because of their handedness.

A genuine massive gravity provides   more physically propagating
modes than the Einstein gravity: 5 and 2 tensor modes. If the
graviton is massive, we expect that they will leave a different
signature on the CMBR anisotropy spectrum.   Similarly, it seems
that there is no way to detect a single  massive graviton directly
by LIGO  even though it has  5 degrees of freedom (DOF). However,
there were some  probes into a  stochastic massive gravitational
wave background   which  is an incoherent superposition of massive
GWs produced by many unsolved astronomical source or by
inflation~\cite{Hayama:2012au,Nishizawa:2013eqa}. In this case, the
observation of  6 polarization modes
($+,~\times,~\circ,~\ell,~x,~y$) is an essential  tool to probe for
the massive gravity.  Here $+$ (plus) and $\times$ (cross) modes are
tensor-type (spin-2) GWs, $\circ$ (breathing) and $\ell$
(longitudinal) are scalar-type (spin-0) GWs. $x$ and $y$ are
vector-type (spin-1) GWs.

In a massive gravity of Einstein-Chern-Simons-Weyl (ECSW) gravity,
however, one has 7[1(scalar)+2(vector)+2(tensor)+2(massive tensor)]
modes~\cite{Stelle:1976gc,Myung:2011nn,Hassan:2013pca} because the
Weyl-squared term could eliminate a longitudinal scalar and the
equation of motion for tensor modes is fourth order.

If the massive graviton exists, its existence could  be proved by
inflation which may play the role of a massive graviton amplifier to
obtain  primordial massive GWs. That is, one method of computing
massive GWs traces their origin to spontaneous emission of single
massive gravitons, which got then  amplified classically by
inflation (expansion) into massive GWs imprinted in the CMBR
temperature and polarization. When one compares massive GWs with
GWs, the difference is twofold: the presence of graviton mass $m$
and number of polarization modes. For the cosmological perturbation
of a massive gravity, one introduces $SO(3)$ decomposition to a
metric tensor which leads to six modes of two scalars, one vector
with 2 modes, and a tensor with 2 under the newtonian gauge.
Usually, the tensor perturbation produces both EE- and BB-mode
polarization power spectra. The vector modes disappear in the
inflationary background of Einstein gravity, while it can be
propagating in the inflationary background  of a massive gravity
theory with Weyl term~\cite{Deruelle:2010kf}.  Two propagating
vector modes reflect that the considering theory belongs to a
massive gravity.  The two scalar modes of $\Phi=-\Psi$ is combined
with the inflaton $\delta\phi$ to give the comoving curvature
perturbation ${\cal R}=-\Phi+(H/\dot{\bar{ \phi}})\delta \phi$ in
the Einstein gravity, while they could become a propagating mode in
the ECSW gravity. However, the Chern-Simons term does not contribute
to the scalar sector because of the parity symmetry in scalar modes.

In this work, we propose  the ECSW gravity theory as a  massive
gravity to detect massive GWs arisen from inflation.  The
cosmological perturbation of Einstein-Weyl gravity has been
performed by showing mainly  that the vector perturbation cannot be
neglected~\cite{Deruelle:2010kf}. This is easily understood when one
recognizes that the Einstein-Weyl gravity describes a massive
gravity with 7 DOF.  We will focus on computing all power spectra by
performing the cosmological perturbations around the de Sitter
inflation. This theory  possesses a tensor ghost as  massive GWs
when one compares with the dRGT massive
gravity~\cite{deRham:2010ik,deRham:2010kj}.

\section{ECSW gravity }

Let us first consider the Einstein-Chern-Simons-Weyl (ECSW) gravity
whose action is given by
\begin{equation} \label{ECSW}
S_{\rm ECSW}=\frac{1}{2\kappa}\int d^4x \sqrt{-g}\Big[ R-2\kappa
\Lambda -\kappa(\partial \phi)^2-\kappa m^2_\phi \phi^2
+\frac{1}{4}\theta~{}^*RR-\frac{1}{2m^2}C^{\mu\nu\rho\sigma}C_{\mu\nu\rho\sigma}\Big],
\end{equation}
where the Chern-Simons term and Weyl-squared term take the form as
\begin{eqnarray}
{}^*RR&=& \frac{1}{2}\epsilon^{\alpha\beta\gamma\delta}R^{\mu\nu}_{~~\alpha\beta}R_{\gamma\delta\mu\nu}, \\
C^{\mu\nu\rho\sigma}C_{\mu\nu\rho\sigma}&=&2\Big(R^{\mu\nu}R_{\mu\nu}-\frac{1}{3}R^2\Big)+
(R^{\mu\nu\rho\sigma}R_{\mu\nu\rho\sigma}-4R^{\mu\nu}R_{\mu\nu}+R^2).
\end{eqnarray}
For our purpose, we include a massive scalar $\phi$ as a
competitor.
 Here we have
$\kappa=8\pi G=1/M^2_{\rm P}$, $M_{\rm P}$ being the reduced Planck
mass. Greek indices run from 0 to 3 with conventions $(-+++)$, while
Latin indices run from 1 to 3. We note that the Chern-Simons term is
coupled to not a scalar $\phi$ but a Chern-Simons scalar $\theta$,
implying the non-dynamical Chern-Simons  gravity
theory~\cite{Jackiw:2003pm,Moon:2014qma}. This contrasts to a
conventional cosmological approach obtained from the dynamical
Chern-Simons gravity with
$f(\phi)~{}^*RR$~\cite{Choi:1999zy,Alexander:2004wk,Saito:2007kt,Satoh:2008ck}.
Further, we note that the Weyl-squared term is invariant under the
conformal transformation of $g_{\mu\nu} \to \Omega^2 g_{\mu\nu}$
like the Maxwell kinetic term of $-F^2/4$ which implies that the
vector and tensor  perturbations are decoupled from the de Sitter
inflation in the limit of $m^2 \to 0 $ (keeping the Weyl-squared
term only).

The Einstein equation is given by
\begin{equation} \label{ein-eq}
G_{\mu\nu}+\kappa \Lambda g_{\mu\nu}+{\cal C}_{\mu\nu}-\frac{1}{m^2}
B_{\mu\nu}= \kappa T_{\mu\nu}
\end{equation} where the Einstein tensor $G_{\mu\nu}$, the Cotton
tensor ${\cal C}_{\mu\nu}$, the Bach tensor $B_{\mu\nu}$, and the
energy-momentum tensor $T_{\mu\nu}$  take the forms
\begin{eqnarray}
G_{\mu\nu}&=&R_{\mu\nu}-\frac{1}{2}Rg_{\mu\nu},\label{einstein} \\
{\cal
C}_{\mu\nu}&=&\nabla_{\gamma}~\theta~\epsilon^{\gamma\rho\sigma}_{~~~(\mu}
\nabla_{|\sigma|}R_{\nu)\rho}+\frac{1}{2}\nabla_{\gamma}\nabla_{\rho}
~\theta~\epsilon_{(\nu}^{~~\gamma\sigma\delta}R^{\rho}_{~~\mu)\sigma\delta},
\label{cottont} \\
 B_{\mu\nu}&=&2
\nabla^\rho\nabla^\sigma
C_{\mu\rho\nu\sigma}+G^{\rho\sigma}C_{\mu\rho\nu\sigma},
\label{bach} \\
T_{\mu\nu}&=&\partial_\mu \phi \partial_\nu
\phi-g_{\mu\nu}\Big(\frac{1}{2} \partial_\rho \phi\partial^\rho
\phi+\frac{m^2_\phi}{2} \phi^2\Big).
\end{eqnarray}
On the other hand, the scalar  equation leads to
\begin{equation} \label{inf-eq}
\nabla^2\phi-m^2_\phi\phi=0,
\end{equation}
while the divergence of the left-hand side of (\ref{ein-eq}) should
be  zero by imposing the Bianchi identity as
\begin{equation} \label{constraint}
\nabla^\mu {\cal C}_{\mu\nu}=-\frac{\nabla_\nu \theta}{8}~{}^*RR=0
\end{equation}
because  $\nabla^\mu T_{\mu\nu}=0$ implies (\ref{inf-eq}) and
$\nabla^\mu B_{\mu\nu}=0$.

 For a conformally flat Friedmann-Robertson-Walker (FRW)
background expressed by the conformal time $\eta$
\begin{eqnarray} \label{frw}
ds^2_{\rm FRW}=a(\eta)^2\Big[-d\eta^2+\delta_{ij}dx^idx^j\Big],
\end{eqnarray}
the Einstein equation and scalar  equation are given by
\begin{eqnarray}
&& \label{cfrw-eq} {\cal H}^2=\frac{\kappa }{3}\Big(a^2\Lambda+\frac{1}{2}(\phi')^2+\frac{a^2}{2}m^2_\phi \phi^2\Big), \\
&&\phi''+2{\cal H}\phi'+a^2m^2_\phi \phi=0,
\end{eqnarray}
where $'$ (prime) denotes differentiation with respect to conformal
time $\eta$ and ${\cal H}=a'/a$.  Here we note that there are no
contributions from the Cotton and Bach tensors because the Cotton
tensor represents a parity-violating term and Bach tensor comes from
a conformal invariant Weyl-squared term, while the FRW universe
preserves the parity symmetry. Also, Eq.(\ref{constraint}) is
satisfied, because
 the  Pontryagin constraint (${}^*RR=0$) is preserved on the FRW background
for $\nabla_\nu\theta\not=0$. This constraint is also   recovered
when one varies the action (\ref{ECSW}) by the field $\theta$.  It
implies that a non-dynamical field $\theta$ remains unfixed in the
background evolution, but it could be determined when one solves the
tensor perturbed equation (see Sec. 3.3)

Now one starts with general perturbed metric
\begin{equation} \label{gpm}
ds^2=a(\eta)^2\Big[-(1+2\Psi)d\eta^2+2B_i d\eta
dx^{i}+(\delta_{ij}+\bar{h}_{ij})dx^idx^j\Big],
\end{equation}
where the SO(3)-decomposition is given by
\begin{eqnarray}
B_i=\partial_iB
+\Psi_i,~\bar{h}_{ij}=2\Phi\delta_{ij}+2\partial_{i}\partial_jE
+\partial_i\bar{E}_j+\partial_j\bar{E}_i+h_{ij}
\end{eqnarray}
with  the transverse vectors $\partial_i\Psi^i=0$,
$\partial_i\bar{E}^i=0$, and transverse-traceless tensor
$\partial_ih^{ij}=h=0$.  To have 7 propagating modes implied by the
massive gravity of ECSW (\ref{ECSW}), we first  choose  the
Newtonian gauge of $B=E=0 $ and $\bar{E}_i=0$ which leads to
12[10+2(massive tensor)]$-$4=8 modes. In this case, the
corresponding perturbed metric and scalar  can be written as
\begin{eqnarray}
ds^2&=&a(\eta)^2\Big[-(1+2\Psi)d\eta^2+2\Psi_i d\eta
dx^{i}+\Big\{(1+2\Phi)\delta_{ij}+h_{ij}\Big\}dx^idx^j\Big],\\
\phi&=&\bar{\phi}+\delta \phi.
\end{eqnarray}
Here $a(\eta)$ and $\bar{\phi}=0$ denote the background spacetime of
de Sitter inflation.  In the case of Einstein gravity, one has a
connection
\begin{equation} \label{pp}
\Psi=-\Phi
\end{equation}
from the linearized  Einstein equation of $\delta
G_i~^j=\partial_i\partial^j(\Psi+\Phi)=0$ and they are not physical
DOF. However, since the ECSW gravity is considered as a massive
gravity with 7 DOF, we  impose one constraint  to meet the massive
gravity with 7 DOF.

 There are two ways to obtain the cosmological
perturbed equations: one is to linearize the Einstein and scalar
equation around the de Sitter inflation  background directly and
the other is first to obtain the bilinear action and then, varying
it to obtain the perturbed equations. In this work, we choose the
second approach.

Now we expand the ECSW action (\ref{ECSW}) up to quadratic order in
the perturbations ($\Psi,\Phi,\delta \phi,\Psi_i,~h_{ij}$) on the de
Sitter background~\cite{Deruelle:2010kf} then the bilinear  actions
for scalar, vector and tensor perturbations can be found as
\begin{eqnarray}
&&\hspace*{-2.3em}\kappa S_{\rm ECSW}^{({\rm S})}=\frac{1}{2}\int
d^4x\Big\{a^2\Big[-6\Phi^{\prime2}+12{\cal
H}\Psi\Phi^{\prime}+2\partial_i\Phi\partial^i\Phi+4\partial_i\Phi\partial^i\Psi
-6{\cal H}^2\Psi^2\nonumber\\
&&\hspace*{8.3em}+\kappa\Big(\delta\phi^{\prime2}
-\partial_i\delta\phi\partial^i\delta\phi -a^2m^2_\phi\delta\phi^2
\Big)\Big] -\frac{2}{3m^2}[(\partial^2 (\Psi-\Phi)]^2\Big\},
\label{scalar}\\
&&\nonumber\\
 &&\hspace*{-2.3em}\kappa S_{\rm ECSW}^{({\rm V})}=\frac{1}{4}\int
d^4x\Big[a^2\partial_i\Psi_j\partial^i\Psi^j
-\theta^{\prime}\epsilon_i^{~jk}\partial^{\ell}\Psi^{i}
\partial_j\partial_{\ell}\Psi_{k}-\frac{1}{m^2}(\partial_i\Psi'_{j}\partial^{i}\Psi'^{j}
-\partial^2\Psi_i\partial^2\Psi^i) \Big],\label{vpeq}\\
&&\nonumber\\
&&\hspace*{-2.3em}\kappa S_{\rm ECSW}^{({\rm T})}=\frac{1}{8}\int
d^4x\Big[a^2(h'_{ij}h'^{ij}-\partial_kh_{ij}\partial^kh^{ij})
-\theta^{\prime}\epsilon_{i}^{~jk}(h'^{\ell i}\partial_jh'_{k\ell}
-\partial^{\ell}h^{pi}\partial_j\partial_{\ell}h_{kp})\nonumber\\
&&\hspace*{10em}-\frac{1}{m^2}(h''_{ij}h''^{ij}
-2\partial_kh'_{ij}\partial^{k}h'^{ij}
+\partial^2h_{ij}\partial^2h^{ij}) \Big]\label{hpeq},
\end{eqnarray}
where $\partial^2\equiv\partial_i\partial^i$ and
$\epsilon^{ijk}\equiv \epsilon^{0ijk}$.

 Varying the actions (\ref{vpeq}) and (\ref{hpeq}) with respect to
$\Psi^{i}$ and $h^{ij}$, respectively leads to the equations of
motion for vector and tensor perturbations as follows:
\begin{eqnarray}
&&\Box\Psi_i-m^2a^2\Psi_i
-m^2\theta^{\prime}\epsilon_i^{~j\ell}\partial_{\ell}\Psi_{j}=0,\label{veq}\\
&&\nonumber\\
&&\Box^2h_{ij}-m^2a^2\Box h_{ij}+2m^2a^2{\cal H}h_{ij}^{\prime}\nonumber\\
&&\hspace*{3em}-~m^2\epsilon_{j}^{~\ell
k}\Big(\theta^{\prime\prime}\partial_{\ell}h_{ki}^{\prime}
+\theta^{\prime}\partial_{\ell}h_{ki}^{\prime\prime}
-\theta^{\prime}\partial_{\ell}\Box h_{ki}\Big)=0\label{heq}.
\end{eqnarray}
 We
emphasize that Eqs.(\ref{veq}) and (\ref{heq}) are newly derived
equations. Turning off the Weyl-squared term (in the limit of $m^2
\to \infty$), Eq.(\ref{veq}) is trivial which implies the
non-propagating vector modes in the modified Chern-Simons
gravity~\cite{Choi:1999zy}. In the other limit of $m^2 \to 0$, we
keep the Weyl-squared term only which is surely independent of $a^2$
because it is invariant under the conformal transformation of
$g_{\mu\nu}\to a^2 \eta_{\mu\nu}$. This implies that the perturbed
field equations  are $\Box\Psi_i=0$ and $\Box^2h_{ij}=0$ which are
independent of the expanding background in the Weyl gravity.

Before we proceed, we briefly mention the scalar perturbations. The
Chern-Simons term does not contribute to the scalar bilinear action
(\ref{scalar}) because of the parity symmetry in scalar modes. Also,
it is important to note that when one compares the last terms in
(\ref{scalar}), (\ref{vpeq}), and (\ref{hpeq}), the last term in
(\ref{scalar}) is a purely (space-like) fourth-order term without
the kinetic term. Hence it could be played the role of a constraint
to reduce 2 DOF to one DOF. In this case, an elegant constraint is
to choose
\begin{equation} \label{phi-psi}
\Psi=\Phi \end{equation}
 which corresponds to taking $m^2\to \infty$
effectively. We note that the constraint (\ref{phi-psi}) is
different from $\Psi=-\Phi$ (\ref{pp}) obtained from the Einstein
gravity. Requiring the condition of $\Psi=\Phi$, the bilinear scalar
action (\ref{scalar}) takes a simple form
\begin{eqnarray} \label{scalar-phi}
\kappa\tilde{S}_{\rm ECSW}^{({\rm S})}=\frac{1}{2}\int
d^4xa^2\Big[-6\Phi^{\prime2}&+&12{\cal
H}\Phi\Phi^{\prime}+6\partial_i\Phi\partial^i\Phi-6{\cal
H}^2\Phi^2\nonumber \\
&+&\kappa\Big(\delta\phi^{\prime2}
-\partial_i\delta\phi\partial^i\delta\phi -a^2m^2_\phi\delta\phi^2
\Big)\Big]
\end{eqnarray}
which is our bilinear scalar action.

\section{Perturbations on de Sitter inflation}

In the de Sitter inflation with constant $H$ and $\bar{\phi}=0$, one
has the background spacetime and Friedmann equation (\ref{cfrw-eq})
\begin{equation}
ds^2_{\rm dS}=-dt^2+e^{2Ht} \delta_{ij} dx^idx^j,~~~H^2=\frac{\kappa
\Lambda }{3}\equiv\frac{m^4_{\rm In}}{3M_{\rm P}^2}
\end{equation}
which implies that
\begin{equation}
a(t)=e^{Ht} \to a(\eta)=-\frac{1}{H\eta}.
\end{equation}
During the De Sitter stage, $a$ goes from a very small to a very
large value which corresponds to $\eta=-\frac{1}{aH}$ running from
$-\infty$ to zero. Also, one has
\begin{equation}  \label{de-con}{\cal H}^2={\cal
H}'=a^2H^2.
\end{equation}
Even though the de Sitter inflation does not provide a graceful exit
when one compares with the slow-roll inflation, we choose  it for a
simple computation.

 The BICEP2 measurement of
 $r=A_T(k_*)/A_s(k_*)=0.2$ together with PLANCK measurement of scalar
 amplitude $A_s=2.215\times 10^{-9}$ determines the scale $m_{\rm I}$
of inflation as~\cite{Krauss:2014sua,Das:2014ada}
 \begin{equation}
 A_T(k_*)=\frac{2}{\pi^2}\Big(\frac{H^2}{M^2_{\rm P}}\Big) \to
 H\simeq 1.1 \times 10^{14} {\rm GeV} \to V^{1/4}=m_{\rm
 I}=2.1\times10^{16} {\rm GeV}
 \end{equation}
 with $M_{\rm P}=2.4 \times 10^{18}$ GeV. This is very close to the
 GUT scale. This implies the small bound
 \begin{equation} \label{hpmass-b}
 \frac{H^2}{M^2_{\rm P}}=2.1\times 10^{-9} \ll1.
\end{equation}

\subsection{Scalar perturbations}

In order to investigate the scalar perturbation in the de Sitter
background, we first derive the linearized scalar equations. Varying
(\ref{scalar-phi}) with respect to $\Phi$ leads to
\begin{equation}
\Phi^{\prime\prime}+2{\cal H}\Phi^{\prime}-4{\cal
H}^2\phi-\partial^2\Phi=0.\label{seq}
\end{equation}
Similarly, varying (\ref{scalar-phi}) with respect to $\delta \phi$
, one has a massive equation
\begin{equation} \label{inflaton-eq1}
\delta \phi'' +2{\cal H} \delta \phi'+m^2_\phi a^2\delta \phi
-\partial^2\delta \phi=0.
\end{equation}
Considering the Fourier expansion of $\Phi$ and $\delta \phi$
\begin{eqnarray}\label{psim}
(\Phi(\eta,{\bf x}),\delta \phi(\eta,{\bf
x}))=\frac{1}{(2\pi)^{\frac{3}{2}}}\int d^3{\bf k}(\Phi_{\bf
k}(\eta),\phi_{\bf k}(\eta))e^{i{\bf k}\cdot{\bf x}},
\end{eqnarray}
the equation (\ref{seq}) can be written as
\begin{eqnarray}\label{s0eq}
\Bigg[\frac{d^2}{d \eta^2}-\frac{2}{\eta}\frac{d}{d
\eta}+k^2-\frac{4}{\eta^2}\Bigg]\Phi_{\bf k}(\eta)=0.
\end{eqnarray}
Now we introduce $z=-\eta k$ and a new variable $v_{\bf
k}=a\Phi_{\bf k}(\eta)\to \frac{k}{H}\frac{1}{z}\Phi_{\bf k}(z)$,
then Eq.(\ref{s0eq}) takes a simple form as
\begin{eqnarray}\label{s1eq}
\Bigg[\frac{d^2}{dz^2}+1-\frac{6}{z^2}\Bigg]v_{\bf k}=0.
\end{eqnarray}
Considering $v_{\bf k}=\sqrt{z}\tilde{v}_{\bf k}$, Eq.(\ref{s1eq})
reduces to the Bessel's equation
\begin{eqnarray}\label{s2eq}
\Bigg[\frac{d^2}{dz^2}+\frac{1}{z}\frac{d}{dz}+1-\frac{\nu^2_\Phi}{z^2}\Bigg]\tilde{v}_{\bf
k}(z)=0
\end{eqnarray}
with
\begin{equation}
\nu_\Phi=\frac{5}{2}.
\end{equation}
The  solution is given by the Hankel function as
\begin{eqnarray}
\Phi_{\bf k}(z)=\frac{\sqrt{z}}{a}H^{(1)}_{5/2}.
\end{eqnarray}

On the other hand, the scalar equation (\ref{inflaton-eq1}) is given
by
\begin{eqnarray}\label{inflaton-eq2}
\Bigg[\frac{d^2}{d \eta^2}-\frac{2}{\eta}\frac{d}{d
\eta}+k^2+\frac{m^2_\phi}{H^2}\frac{1}{\eta^2}\Bigg]\phi_{\bf
k}(\eta)=0,
\end{eqnarray}
which can be further transformed into
\begin{eqnarray}\label{inflaton-eq3}
\Bigg[\frac{d^2}{d\eta^2}+k^2-\frac{2}{\eta^2}+\frac{m^2_\phi}{H^2}\frac{1}{\eta^2}\Bigg]\tilde{\phi}_{\bf
k}(\eta)=0
\end{eqnarray}
for $\tilde{\phi}_{\bf k}=a\phi_{\bf k}=-\phi_{\bf k}/(H\eta)$.
After expressing (\ref{inflaton-eq3}) in terms of $z=-k\eta$ and
then introducing $\tilde{\phi}_{\bf
k}=\sqrt{z}\tilde{\tilde{\phi}}_{\bf k}$,  it  leads to the Bessel's
equation as
\begin{eqnarray}\label{inflaton-eq4}
\Bigg[\frac{d^2}{dz^2}+\frac{1}{z}\frac{d}{dz}+1-\frac{\nu^2_\phi}{z^2}\Bigg]\tilde{\tilde{\phi}}_{\bf
k}(z)=0
\end{eqnarray}
with
\begin{equation}
\nu_\phi=\sqrt{\frac{9}{4}-\frac{m^2_\phi}{H^2}}.
\end{equation}
The solution to (\ref{inflaton-eq4}) is given by the Hankel function
$H^{(1)}_\nu$. Accordingly, one has the solution to
(\ref{inflaton-eq2})
\begin{equation} \label{inflaton-eq5}
\phi_{\bf k}(z)=\frac{\sqrt{z}}{a}\tilde{\tilde{\phi}}_{\bf
k}=\frac{\sqrt{z}}{a}H^{(1)}_{\nu_\phi}(z).
\end{equation}

\subsection{Vector perturbations}

We first consider Eq.(\ref{veq}) for vector perturbation and expand
the mode $\Psi_i$ in plane waves with the right-handed and
left-handed circularly polarized states
\begin{eqnarray}\label{psim}
\Psi_i(\eta,{\bf x})=\frac{1}{(2\pi)^{\frac{3}{2}}}\int d^3{\bf
k}\sum_{s={\rm R,L}}\tilde{p}_i^{s}({\bf k})\Psi_{\bf
k}^{s}(\eta)e^{i{\bf k}\cdot{\bf x}},
\end{eqnarray}
where $\tilde{p}_i^{s}$ are  the polarization vectors for the
right-handed or left-handed circularly polarized state.  They are
defined by $\tilde{p}^{R/L}_{i}=\frac{1}{\sqrt{2}}(p^1_{i}\pm i p^2
_{i})$ with $p^{1/2}_{i}$ linear polarization   vectors. We note
that $\tilde{p}^{R/L}_i (\tilde{p}^{R/L, i})^*=1$, while $p^{1/2}_i
p^{1/2, i}=1$.  Also, circularly polarized vector mode $\Psi_{\bf
k}^{s}$ are defined by $\Psi_{\bf k}^{s}=\frac{1}{\sqrt{2}}(v_{\bf
k}^{1}\mp i v_{\bf k}^{2})$ with $v_{\bf k}^{1/2}$ linearly
polarized vector modes.

Plugging (\ref{psim}) into the equation (\ref{veq}), one finds the
equation
\begin{eqnarray}\label{v0eq}
\Bigg[\frac{d^2}{d\eta^2}+k^2+m^2\Big(\frac{1}{\eta^2H^2}
-\lambda^sk\frac{d\theta}{d\eta}\Big)\Bigg]\Psi_{\bf k}^s(\eta)=0,
\end{eqnarray}
where $\lambda^{\rm R/L}=\pm1$.  In deriving Eq.(\ref{v0eq}),  the
following relation was used:
\begin{eqnarray}
ik_c\epsilon_{a}^{~cd}\tilde{p}_{d}^{s}=\lambda^{s}
k\tilde{p}_{a}^{s}.
\end{eqnarray}
At this stage, we choose a non-dynamical field $\theta$ to solve
(\ref{v0eq}). We mention that $\theta$ remains unfixed in the
background evolution, but it must be determined when one tries  to
solve the vector perturbed equation (\ref{v0eq}). Note that in the
inflation model with the Gauss-Bonnet and the parity violating
corrections, it is given by $\theta=c\ln \eta$~\cite{Satoh:2008ck}
to have slow-roll inflation, while it will take the form
$\theta=c/\eta$ in the ECSW gravity to make factorization of
fourth-order tensor equation (see Sec.3.3). Now we choose
$\theta=c\ln \eta$, then one has $\theta'=c/\eta$. In this case,
Eq.(\ref{v0eq}) takes the form
\begin{equation}\label{v1eq}
\Bigg[\frac{d^2}{d\eta^2}+k^2+m^2\Big(\frac{1}{\eta^2H^2}
-\frac{\lambda^sk c}{\eta}\Big)\Bigg]\Psi_{\bf k}^s(\eta)=0,
\end{equation}
which could describe a propagation of circularly polarized vector
waves.

 For $\theta=c/\eta$, however, one has
 $\theta'=-c/\eta^2$. In this case, Eq.(\ref{v0eq}) reduces to
\begin{equation}\label{v2eq}
\Bigg[\frac{d^2}{d\eta^2}+k^2+m^2\Big(\frac{1}{\eta^2H^2}
+\frac{\lambda^skc}{\eta^2}\Big)\Bigg]\Psi_{\bf k}^s(\eta)=0,
\end{equation}
which is the same as  the massive tensor equation (\ref{hcsw}).  The
above shows that the vector equation  depends on the choice of
$\theta$.

Finally, for $\theta=0$,  Eq.(\ref{v0eq}) reduces to
\begin{equation}\label{v2eq}
\Big[\frac{d^2}{d\eta^2}+k^2+\frac{m^2}{\eta^2H^2} \Big]\Psi_{\bf
k}(\eta)=0,
\end{equation}
which is just the massive equation for the transverse vector ${\cal
A}^\bot$~\cite{Dimopoulos:2006ms}, while the equation of
longitudinal vector ${\cal A}^\|$ leads to  the scalar equation
(\ref{inflaton-eq3}) when we consider the massive Maxwell Lagrangian
after plugging $g_{\mu\nu}=a^2 \eta_{\mu\nu}$
\begin{equation} \label{maxwell}
{\cal
L}_{M}=\Big(-\frac{1}{4}F^{\mu\nu}F_{\mu\nu}+\frac{m_{F}^2}{2} a^2
A^\mu A_\mu\Big).
\end{equation}
Here the first term preserves conformal symmetry like the
Weyl-squared term, while the second term breaks the conformal
symmetry. In the limit of $m_{F}^2 \to 0$, one recovers the
conformally invariant Maxwell term.

\subsection{Tensor perturbations}

Now we turn to the equation (\ref{heq}) for tensor perturbations. In
this case, the metric tensor $h_{ij}$ can be expanded in Fourier
modes
\begin{eqnarray}\label{hijm}
h_{ij}(\eta,{\bf x})=\frac{1}{(2\pi)^{\frac{3}{2}}}\int d^3{\bf
k}\sum_{s={\rm R,L}}\tilde{p}_{ij}^{s}({\bf k})h_{\bf
k}^{s}(\eta)e^{i{\bf k}\cdot{\bf x}},
\end{eqnarray}
where $\tilde{p}_{ij}^{s}$ are  the polarization tensors for the
right-handed or left-handed circularly polarized state. They are
defined by $\tilde{p}^{R/L}_{ij}=\frac{1}{\sqrt{2}}(p^+_{ij}\pm i
p^\times _{ij})$ with $p^{+/\times}_{ij}$ linear polarization
tensors.  We note that $\tilde{p}^{R/L}_{ij}( \tilde{p}^{R/L,
ij})^*=1$, while $p^{+/\times}_{ij} p^{+/\times, ij}=1$. Also,
circularly polarized tensor mode $h_{\bf k}^{s}$ is  defined by
$h_{\bf k}^{s}=\frac{1}{\sqrt{2}}(h_{\bf k}^{+}\mp i h_{\bf
k}^{\times})$ with $h_{\bf k}^{+/\times}$ linearly polarized tensor
modes.

Plugging (\ref{hijm}) into (\ref{heq}) leads to the fourth-order
differential equation
\begin{eqnarray}
&&(h_{\bf k}^{s})^{''''}+2k^2(h_{\bf k}^{s})^{''}+k^4h_{\bf k}^{s}
+m^2\Big(a^2-\lambda^sk\theta^{\prime}\Big)(h_{\bf k}^{s})^{''}
\nonumber\\
&&\hspace*{7em}+~m^2\Big(2a^2{\cal
H}-\lambda^sk\theta^{\prime\prime}\Big)(h_{\bf
k}^{s})^{'}+m^2\Big(a^2-\lambda^sk\theta^{\prime}\Big)k^2h_{\bf
k}^{s}=0,\label{heq2}
\end{eqnarray}
where we used
\begin{eqnarray}
ik_c\epsilon_{a}^{~cd}\tilde{p}_{bd}^{\rm
s}=\lambda^{s}k\tilde{p}_{ab}^{\rm s}.
\end{eqnarray}
It is important  to note that factorizing the fourth-order equation
(\ref{heq2}) into two second-order equations is a nontrivial task
because the Chern-Simons field $\theta$ and its derivatives are
present.

 In the limit of $m^2\to \infty$, one recovers the tensor
perturbation equation for the Chern-Simons modified gravity which is
surely a second-order equation~\cite{Alexander:2004wk}
\begin{eqnarray}
\Big(1-\lambda^sk\frac{\theta^{\prime}}{a^2}\Big)(h_{\bf
k}^{s,CS})^{''} +\Big(2{\cal
H}-\lambda^sk\frac{\theta^{\prime\prime}}{a^2}\Big)(h_{\bf
k}^{s,CS})^{'}+\Big(1-\lambda^sk\frac{\theta^{\prime}}{a^2}\Big)k^2h_{\bf
k}^{s,CS}=0,\label{cs-heq2}
\end{eqnarray}
which is transformed into the Mukhanov-Sassaki type equation
\begin{equation}\label{ms-eq}
(\mu_{\bf k}^{s,CS})''+\Big(k^2-\frac{z_s''}{z_s}\Big)\mu_{\bf
k}^{s,CS}=0
\end{equation}
when one introduces
\begin{equation}
z_s(\eta,k)=a\sqrt{1-\lambda^sk \frac{\theta'}{a^2}},~~\mu_{\bf
k}^{s,CS}=z_s h_{\bf k}^{s,CS}.
\end{equation}
The effective potential $z''/z$ depends not only on time $\eta$ and
polarization $\lambda^s$, but also on the wave number $k$, which
shows a  difference when comparing  with the standard case
$z(\eta)$~\cite{Saito:2007kt}. For $\theta=c\ln \eta$, one has
$\theta'/a^2=cH^2\eta$, where
$c=-\Omega/(M_cH)$~\cite{Satoh:2008ck}.  In this case,
Eq.(\ref{ms-eq}) takes the form
\begin{equation}\label{slowms-eq}
(\mu_{\bf k}^{s,CS})''+\Big(k^2-\frac{2}{\eta^2}+\frac{\lambda_s
kH\Omega}{M_c}\frac{1}{\eta}\Big)\mu_{\bf k}^{s,CS}(\eta)=0,
\end{equation}
which could describe a propagation of circularly polarized GWs. Here
$\Omega=M_c \dot{\theta}/(2M^2_{\rm P})$ was considered to be a
nearly constant with $M_c=k/a$.

 For $\theta=c/\eta$, however, one has
 $\theta'/a^2=-cH^2$ which implies
$z_s=a\sqrt{1+\lambda_s k cH^2}$. Then, Eq.(\ref{ms-eq}) reduces to
\begin{equation}\label{sms-eq}
(\mu_{\bf k}^{s,CS})''+\Big(k^2-\frac{2}{\eta^2}\Big)\mu_{\bf
k}^{s,CS}(\eta)=0
\end{equation}
which is just the tensor perturbation equation (\ref{hee}). The
above shows that the tensor perturbed equation  depends on the
choice of $\theta$.

 However, it is shown that the
Eq.(\ref{heq2}) can be factorized as the following two different
types (see Appendix):
\begin{eqnarray}
&&\hspace*{-3em}\Bigg[\frac{d^2}{d\eta^2}+\frac{2}{\eta}\frac{d}{d\eta}
+k^2+m^2\Big(\frac{1}{\eta^2H^2} -\lambda^s
k\frac{d\theta}{d\eta}\Big)\Bigg]
\Bigg[\frac{d^2}{d\eta^2}-\frac{2}{\eta}\frac{d}{d\eta}
+k^2\Bigg]h_{\bf k}^{s}=0,\label{dec1}\\
&&\nonumber\\
&&\hspace*{-3em}\Bigg[\frac{d^2}{d\eta^2}-\frac{2}{\eta}\frac{d}{d\eta}
+k^2\Bigg] \Bigg[\eta^2\frac{d^2}{d\eta^2}-2\eta\frac{d}{d\eta}
+2+k^2\eta^2+m^2\Big(\frac{1}{H^2}
-\lambda^sk\eta^2\frac{d\theta}{d\eta}\Big)\Bigg]h_{\bf
k}^{s}=0\label{dec2}
\end{eqnarray}
when one chooses \begin{equation}
\label{theta}\theta=c_1+\frac{c_2}{\eta}
\end{equation}
 with integration constants $c_1$ and $c_2$. Their  mass
dimensions of $c_1$ and $c_2$ are given by  $[{\rm M}]^{-2}$ and
$[{\rm M}]^{-3}$, respectively. The choice of (\ref{theta})
contrasts to the  dynamical Chern-Simons coupling studied in
~\cite{Choi:1999zy,Alexander:2004wk,Saito:2007kt,Satoh:2008ck}.

Introducing a new quantity $\mu_{\bf k}^{s}$ defined by $h_{\bf
k}^{s}=\eta \mu_{\bf k}^{s}$, one can read off the Einstein (E) and
Chern-Simons-Weyl (CSW) mode equations  from Eqs.(\ref{dec1}) and
(\ref{dec2}) as
\begin{eqnarray}
&&\Bigg[\frac{d^2}{d\eta^2}+k^2-\frac{2}{\eta^2}\Bigg]\mu_{\bf
k}^{s({\rm E})}=0,
\label{hee}\\
&&\nonumber\\
&&\Bigg[\frac{d^2}{d\eta^2}+k^2+m^2\Big(\frac{1}{\eta^2H^2}
-\lambda^sk\frac{d\theta}{d\eta}\Big)\Bigg]\mu_{\bf k}^{s({\rm
CSW})}=0\label{hcsw}.
\end{eqnarray}
We note that Eq.(\ref{hcsw}) is exactly the same form  as
Eq.(\ref{v0eq}), obtained for the vector perturbation $\Psi_{\bf
k}^s$ where $\theta$ is undetermined. For $c_1=0$ and $c_2=M_{\rm
P}^{-3}$ together with $z=-k\eta$~\cite{Alexander:2004wk}, Eqs.
(\ref{hee}) and (\ref{hcsw}) take the forms
\begin{eqnarray}\label{hc0}
&&\Big[\frac{d^2}{dz^2}+1-\frac{2}{z^2} \Big]\mu_{\bf k}^{s({\rm
E})}=0, \\
&&\label{hc11}
\Big[\frac{d^2}{dz^2}+1+\frac{\tilde{m}_s^2}{H^2}\frac{1}{z^2}
\Big]\mu_{\bf k}^{s({\rm CSW})}=0
\end{eqnarray}
with
 \begin{equation}
 \tilde{m}^2_s=m^2(1+\lambda^skH^2M_{\rm
P}^{-3}).
\end{equation}

It is easy to show that the tensor solution to (\ref{hc0}) is given
by
\begin{equation} \label{emode}
\mu_{\bf k}^{s({\rm E})}=\alpha e^{iz}\Big(1+\frac{i}{z}\Big)+\beta
e^{-iz}\Big(1-\frac{i}{z}\Big),
\end{equation}
where $\alpha$ and $\beta$ are the undetermined normalization
constants.

  Introducing
\begin{equation}
\mu_{\bf k}^{s({\rm CSW})}=\sqrt{z}\tilde{\mu}_{\bf k}^{s({\rm
CSW})},
\end{equation}
Eq.(\ref{hc11}) reduces to the Bessel equation
\begin{equation} \label{hc2}
\Big[\frac{d^2}{dz^2}+ \frac{1}{z}\frac{d}{dz}+1-\frac{\nu^2_s}{z^2}
\Big]\tilde{\mu}_{\bf k}^{s({\rm CSW})}=0
\end{equation}
whose solution is given by the Hankel function
\begin{equation}\label{ns}
\tilde{\mu}_{\bf k}^{s({\rm
CSW})}=H^{(1)}_{\nu_s}(z),~~\nu_s=\sqrt{\frac{1}{4}-\frac{\tilde{m}_s^2}{H^2}}<\frac{1}{2}.
\end{equation}
Considering the bound (\ref{hpmass-b}), we expect to have
$kH^2/M_{\rm P}^3 \ll 1$. It means that we can treat the
parity-violating effect as a small correction in Eq. (\ref{hc11}).
Here, requiring that the index $\nu_s$  be positive leads to the
condition
\begin{equation} \label{mass-b}
\frac{\tilde{m}_s^2}{H^2}<\frac{1}{4} \to m^2<\frac{H^2}{4}\to m
<5.5\times 10^{13} {\rm GeV}
\end{equation}
which corresponds to the graviton mass bound.

As a byproduct, if the Einstein-mode equation (\ref{hc0}) is
expressed in terms of the Bessel equation, it gives $\nu_s=3/2$. Its
solution is found to be
\begin{equation}
\mu_{\bf k}^{s({\rm
E})}=\sqrt{z}H^{(1)}_{3/2}(z)=\sqrt{\frac{2}{\pi}}
e^{-i\pi}e^{iz}\Big(1+\frac{i}{z}\Big)
\end{equation}
which is the first term of (\ref{emode}), while the second term is
given by $\sqrt{z}H^{(2)}_{3/2}(z)$.

Finally,  the two  tensor modes are given by
\begin{equation} \label{n-tensor}
\Big\{ h_{\bf k}^{s({\rm E})}(z),h_{\bf k}^{s({\rm
CSW})}(z)\Big\}=\Big\{z^{\frac{3}{2}}H^{(1)}_{3/2}(z),z^{\frac{3}{2}}H^{(1)}_{\nu_s}(z)\Big\}.
\end{equation}
For later convenience, we list asymptotic forms of the Hankel
function
\begin{equation} \label{hankela}
H^{(1)}_{\nu_s}(z)\Big|_{z\to \infty} \sim \sqrt{\frac{2}{\pi z}}e^{
i(z-\frac{\nu_s \pi}{2}-\frac{\pi}{4})},~~H^{(1)}_{\nu_s}\Big|_{z\to
0} \sim \frac{i \Gamma(\nu_s)}{\pi}\frac{1}{(\frac{z}{2})^{\nu_s}}.
\end{equation}

\section{Primordial power spectra}

The power spectrum is usually given by the two-point correlation
function which is calculated in the vacuum state $|0>$. It is
defined by
\begin{equation}
<0|{\cal F}(\eta,\bold{x}){\cal F}(\eta,\bold{x}')|0>=\int
d^3\bold{k} \frac{{\cal P}_{\cal F}}{4\pi k^3}e^{-i \bold{k}\cdot
(\bold{x}-\bold{x}')},
\end{equation}
where ${\cal F}$ denotes a scalar, vector, and tensor and
$k=|\bold{k}|$ is the wave number. Fluctuations are created on all
length scales with  wave number $k$. Cosmologically relevant
fluctuations start their lives inside the Hubble radius which
defines the subhorizon as
\begin{equation}
{\rm subhorizon}:~~k~\gg aH~(z=-k\eta\gg 1).
\end{equation}
On the other hand, the comoving Hubble radius $(aH)^{-1}$ shrinks
during inflation while the comoving wavenumber $k$ is constant.
Therefore, eventually all fluctuations exit the comoving  Hubble
radius which defines the superhorizon as
\begin{equation}
{\rm superhorizon}:~~k~\ll aH~(z=-k\eta\ll 1).
\end{equation}
We might calculate the quantum-mechanical variance of fluctuations
(two-point function)  by taking the Bunch-Davies vacuum $|0>$ in the
de Sitter inflation. In the de Sitter inflation, we choose the limit
of  $z\to \infty$ (subhorizon) to define the Bunch-Davies vacuum,
while we choose the limit of $z\to 0$ to obtain a definite form of
power spectra.

In general, all fluctuations of scalar and tensor originate on
subhorizon scales and they propagate for a long time on superhorizon
scales. This can be checked by computing their power spectra which
are scale-invariant. However, it would be interesting  to check what
happens when one computes the power spectra of  the massive
fluctuations.

\subsection{Scalar  power spectra}

In this section, we first calculate scalar power spectrum. To this
end, we consider the conjugate momentum for the field $\Phi$, which
is defined by
\begin{eqnarray}\label{sconj}
\pi_{\Phi}=\frac{6a^2}{\kappa}{\Phi}^{\prime},
\end{eqnarray}
being obtained from the scalar action (\ref{scalar-phi}) in the de
Sitter background. The canonical quantization is implemented by
imposing commutation relation
\begin{eqnarray}\label{scomm}
[\hat{\Phi}(\eta,{\bf x}),\hat{\pi}_{\Phi}(\eta,{\bf
x}^{\prime})]=i\delta({\bf x}-{\bf x}^{\prime})
\end{eqnarray}
with $\hbar=1$. Now, the operator $\hat{\Phi}$ can be expanded in
Fourier modes as
\begin{eqnarray}\label{phie}
\hat{\Phi}(\eta,{\bf x})=\frac{1}{(2\pi)^{\frac{3}{2}}}\int d^3{\bf
k}\Big(\hat{a}_{\bf k}\Phi_{\bf k}(\eta)e^{i{\bf k}\cdot{\bf
x}}+h.c.\Big)
\end{eqnarray}
and the operator $\hat{\pi}_{\Phi}=\frac{6a^2}{\kappa}\Phi^{\prime}$
is given by (\ref{phie}). Substitution of (\ref{phie}) and
$\hat{\pi}_{\Phi}$ into (\ref{scomm}) leads to the commutation
relation and Wronskian condition as
\begin{eqnarray}
&&\hspace*{-2em}[\hat{a}_{\bf k},\hat{a}_{{\bf
k}^{\prime}}^{\dag}]=\delta({\bf k}-{\bf
k}^{\prime}),\label{comm0}\\
\nonumber\\
&&\hspace*{-2em}\Phi_{\bf k}\Big(\frac{6a^2}{\kappa}\Big)(\Phi_{\bf
k}^{*})^{\prime}-c.c.=i \to \Phi_{\bf k}\frac{d\Phi_{\bf
k}^{*}}{dz}-c.c.=-\frac{2i\kappa}{12ka^2}. \label{swcon}
\end{eqnarray}
A next step is to choose the initial mode solution to define the
Bunch-Davies vacuum $|0>$ when $z\to\infty$. We note that
 the solution of $v_{\bf k}=a\Phi_{\bf k}$ is
 given to be
\begin{eqnarray}\label{vinf}
v_{{\bf k},z\to\infty}\sim e^{iz},
\end{eqnarray}
as a solution to the asymptotic scalar equation
\begin{eqnarray}\label{s0eq1}
\Big[\frac{d^2}{dz^2}+1\Big]v_{{\bf k},z\to \infty}(z)=0,
\end{eqnarray}
which implies the normalized solution
\begin{equation}\label{s-sol11} \Phi_{{\bf k},z\to \infty}\sim
\frac{H}{\sqrt{12k^3}}ze^{iz}.
\end{equation}
This is  a plane wave to define the Bunch-Davies vacuum.
On the
other hand, in the superhorizon limit of $z\to 0$, one has  a
solution
\begin{eqnarray}\label{vinf}
v_{{\bf k},z\to 0}\sim \frac{1}{z^2},
\end{eqnarray}
as a solution to
\begin{eqnarray}\label{s0eq2}
\Big[\frac{d^2}{dz^2}-\frac{6}{z^2}\Big]v_{{\bf k},z\to 0}(z)=0.
\end{eqnarray}
It implies that
\begin{equation}\label{s-sol12} \Phi_{{\bf k},z\to
0}\sim \frac{1}{z}
\end{equation}
which means that $\Phi_{\bf k}$ diverges   in  the superhorizon
limit. Then, the power spectrum is given  by
\begin{equation} \label{msp0}
{\cal P}_{\Phi}=\frac{k^3}{2\pi^2} |\Phi_{\bf
k}|^2=\frac{H^2}{48\pi}z^3|e^{i\frac{3\pi }{2}}H_{5/2}^{(1)}(z)|^2.
\end{equation}
In the superhorizon limit of $z\to 0$,   the scalar power spectrum
is given by
\begin{eqnarray}
{\cal P}_{\Phi}\Big|_{z\to 0} \sim
\frac{1}{6}\Big(\frac{H}{2\pi}\Big)^2\Big[\frac{\Gamma(5/2)}{\Gamma(3/2)}\Big]^2\Big(\frac{k}{2aH}\Big)^{-2}
\sim \frac{1}{z^2}
\end{eqnarray}
which blows up in the superhorizon limit of $z\to 0$.

To obtain the power spectrum for a massive scalar $\delta \phi$, we
obtain the conjugate momentum
\begin{equation} \pi_{\delta\phi}=a^2 \delta \phi'
\end{equation}
which defines the commutator \begin{equation} [\delta
\hat{\phi}(\eta, {\bf x}),  \hat{\pi}_{\delta\phi}(\eta, {\bf x}')
]=i\delta({\bf x},{\bf x}'). \end{equation} This implies that the
Wronskian condition and commutator are given by
\begin{equation} \label{s-wron}
a^2\Big(\phi_{\bf k}(\phi^*_{\bf k})'-\phi^*_{\bf k}(\phi_{\bf
k})'\Big)=i,~~[\hat{a}_{\bf k},\hat{a}^\dagger_{\bf
k'}]=\delta({\bf k}-{\bf k}').
\end{equation}
We note that the Wronskian condition  together
with(\ref{inflaton-eq5}) determines the normalized scalar mode
\begin{equation}
\label{infla4} \phi_{\bf
k}(z)=\frac{1}{\sqrt{2}}\sqrt{\frac{\pi}{2}} e^{i(\frac{\pi
\nu_\phi}{2}+\frac{\pi}{4})}\frac{\sqrt{-\eta}}{a}H_{\nu_\phi}^{(1)}(z),
\end{equation}
where the first factor of $1/\sqrt{2}$ is the normalization from
the Wronskian condition (\ref{s-wron}) as
$i=2i\times(1/\sqrt{2})^2$. Then, the power spectrum is defined by
\begin{equation} \label{msp0}
{\cal P}_{\delta\phi}=\frac{k^3}{2\pi^2} |\phi_{\bf
k}|^2=\frac{H^2}{8\pi}z^3|e^{i(\frac{\pi
\nu_\phi}{2}+\frac{\pi}{4})}H_{\nu_\phi}^{(1)}(z)|^2.
\end{equation}
In the case of $\nu_\phi=3/2(m^2_\phi=0)$, it leads to the power
spectrum for a massless scalar  as
\begin{eqnarray}
{\cal P}_{\delta\phi}\Big|_{m^2_{\phi} \to 0,z\to 0} \sim
\Big(\frac{H}{2\pi}\Big)^2
\end{eqnarray}
which is a scale-invariant spectrum.

 In the limit of $z\to 0$,  one refers to the form
(\ref{hankela}) which implies that the superhorizon limit of the
inflaton  power spectrum is given by
\begin{eqnarray}
{\cal P}_{\delta\phi}\Big|_{z\to 0} \sim
\Big(\frac{H}{2\pi}\Big)^2\Big[\frac{\Gamma(\nu_\phi)}{\Gamma(3/2)}\Big]^2\Big(\frac{k}{2aH}\Big)^{3-2\nu_\phi}.
\end{eqnarray}
Assuming  $m^2_\phi/H^2\ll 1$ such that $\nu_\phi \simeq
3/2-m^2_\phi/(3H^2)+{\cal O}(m^4_\phi/H^4)$, one has
\begin{equation}
{\cal
P}_{\delta\phi}\simeq\Big(\frac{H}{2\pi}\Big)^2\Big(\frac{k}{2aH}\Big)^{\frac{2m^2_\phi}{3H^2}}
\end{equation}
whose spectral index takes the form
\begin{equation}
n_{\delta \phi}-1=\frac{d\ln{\cal P}_{\delta\phi}}{d\ln k}\simeq
\frac{2m^2_\phi}{3H^2}.
\end{equation}
In the case of a finite mass, the spectrum  would be slightly blue
due to the massive nature. However, for $m^2_\phi\ll H^2$, the
spectrum is almost scale-invariant and the condition of $m^2_\phi
\ll H^2$ determines a long-lasting de Sitter inflation.

Since a longitudinally light massive vector ${\cal A}^\|$ satisfies
the massive scalar equation (\ref{inflaton-eq3}), we expect to have
its power spectrum as
\begin{equation}
{\cal P}_{{\cal A}^\|}=\frac{k^3}{2\pi^2} |{\cal A}^\|_{\bf
k}|^2=\frac{H^2}{8\pi}z^3|e^{i(\frac{\pi \nu_{{\cal
A}^\|}}{2}+\frac{\pi}{4})}H_{\nu_{{\cal A}^\|}}^{(1)}(z)|^2
\end{equation}
with
\begin{equation}
\nu_{{\cal A}^\|}=\sqrt{\frac{9}{4}-\frac{m^2_F}{H^2}}.
\end{equation}
Considering  $m^2_F/H^2\ll 1$ such that $\nu_F \simeq
3/2-m^2_F/(3H^2)+{\cal O}(m^4_F/H^4)$, we have
\begin{equation}
{\cal P}_{{\cal
A}^\|}\simeq\Big(\frac{H}{2\pi}\Big)^2\Big(\frac{k}{2aH}\Big)^{\frac{2m^2_F}{3H^2}}
\end{equation}
whose spectral index takes the form~\cite{Dimopoulos:2006ms}
\begin{equation}
n_{{\cal A}^\|}-1=\frac{d\ln{\cal P}_{{\cal A}^\|}}{d\ln k}\simeq
\frac{2m^2_F}{3H^2}.
\end{equation}

\subsection{Vector  power spectrum}

We now calculate vector power spectrum. For this purpose, we define
a commutation relation for the vector. In the bilinear  action
(\ref{vpeq}), the conjugate momentum for the field $\Psi_j$ is
defined by
\begin{eqnarray}\label{vconj}
\pi_{\Psi}^{j}=\frac{1}{2\kappa m^2}\Psi^{j\prime}.
\end{eqnarray}
The canonical quantization is implemented  by imposing the
commutation relation
\begin{eqnarray}\label{vcomm}
[\hat{\Psi}_{j}(\eta,{\bf x}),\hat{\pi}_{\Psi}^{j}(\eta,{\bf
x}^{\prime})]=2i\delta({\bf x}-{\bf x}^{\prime})
\end{eqnarray}
with $\hbar=1$.

Now, the operator $\hat{\Psi}_{j}$ can be expanded in Fourier modes
as
\begin{eqnarray}\label{vex}
\hat{\Psi}_{j}(\eta,{\bf x})=\frac{1}{(2\pi)^{\frac{3}{2}}}\int
d^3{\bf k}\sum_{s={\rm R,L}}\Big(\tilde{p}_{j}^{s}({\bf
k})\hat{a}_{\bf k}^{s}\Psi_{\bf k}^{s}(\eta)e^{i{\bf k}\cdot{\bf
x}}+h.c.\Big)
\end{eqnarray}
and the operator $\hat{\pi}_{\Psi}^{j}=\frac{1}{2\kappa
m^2}\hat{\Psi}^{j\prime}$ can be easily obtained from (\ref{vex}).
Here the circularly polarized vector $\tilde{p}_{j}^{\rm L}$
satisfies $\tilde{p}_{j}^{\rm L}(\tilde{p}^{j\rm
L})^*=\tilde{p}_{j}^{\rm R}(\tilde{p}^{j\rm R})^*=1$ and the
superscript $s$ in (\ref{vex}) denotes  $({\rm L,R})$ circularly
polarized vector.

Plugging (\ref{vex}) and $\hat{\pi}_{\Psi}^{j}$ into (\ref{vcomm}),
we find the commutation relation and Wronskian condition as
\begin{eqnarray}
&&\hspace*{-2em}[\hat{a}_{\bf k}^{s},\hat{a}_{\bf k^{\prime}}^{
s^{\prime}\dag}]=\delta^{ss^{\prime}}\delta({\bf k}-{\bf
k}^{\prime}),\label{comm0}\\
\\
&&\hspace*{-2em}\Psi_{\bf k}^{s}\Big(-\frac{2}{\kappa
m^2}\Big)(\Psi_{\bf k}^{*s})^{\prime}-c.c.=-4i \to \Psi_{\bf
k}^{s}\frac{d\Psi_{\bf k}^{*s}}{dz}-c.c.=-\frac{2i\kappa m^2}{k}.
\label{vwcon}
\end{eqnarray}
  We
choose the initial mode solution for a Bunch-Davies vacuum $|0>$
\begin{eqnarray}
\Psi_{{\bf k},z\to \infty}^{s} \sim\sqrt{\frac{\kappa m^2}{k}}
~\underbrace{e^{iz}}=\sqrt{\frac{\kappa
m^2}{k}}\underbrace{\sqrt{\frac{\pi}{2}}e^{i(\frac{\pi
\nu_s}{2}+\frac{\pi}{4})}\sqrt{z}H^{(1)}_{\nu_s}(z\to \infty)}
\end{eqnarray}
which is obtained as a solution to the asymptotic vector equation
\begin{eqnarray}\label{v0eq1}
\Big[\frac{d^2}{dz^2}+1\Big]\Psi_{{\bf k},z\to \infty}^s(z)=0
\end{eqnarray}
together with the Wronskian condition (\ref{vwcon}). In this case,
Eq. (\ref{v0eq}) can be written as
\begin{eqnarray}\label{v0eq2}
\Big[\frac{d^2}{dz^2}+1+\frac{\tilde{m}^2_s}{H^2}\frac{1}{z^2}\Big]\Psi_{\bf
k}^s(z)=0.
\end{eqnarray}
for $\theta=(M_{\rm p}^{3}\eta)^{-1}$. The full solution to
(\ref{v0eq2}) is given by the Hankel function
\begin{eqnarray}\label{vsol2}
\Psi_{\bf k}^{s}&=&\sqrt{\frac{\kappa m^2}{k}}
\sqrt{\frac{\pi}{2}}e^{i(\frac{\pi
\nu_s}{2}+\frac{\pi}{4})}\sqrt{z}H_{\nu_s}^{(1)}(z),\label{s2f}
\end{eqnarray}
where  $\nu_s<1/2$ is given by  (\ref{ns}). On the other hand, the
vector power spectrum is defined by
\begin{eqnarray}\label{powerv}
\langle0|\hat{\Psi}_{j}(\eta,{\bf x})\hat{\Psi}^{j}(\eta,{\bf
x})|0\rangle=\int d^3{\bf k}\frac{{\cal P}_{\Psi}}{4\pi
k^3}e^{i{\bf k}\cdot({\bf x}-{\bf x^{\prime}})},
\end{eqnarray}
where we used the Bunch-Davies vacuum state imposing $\hat{a}_{\bf
k}^{s}|0\rangle=0$ and a quantity ${\cal P}_{\Psi}$ in
(\ref{powerv}) denotes ${\cal P}_{\Psi}\equiv\sum_{s={\rm
R,L}}\frac{k^3}{2\pi^2}\Big|\Psi_{\bf k}^{s}\Big|^2$. Plugging
(\ref{vsol2}) into (\ref{powerv}), we find the scale-dependent
spectrum
\begin{eqnarray}
{\cal P}_{\Psi}=\sum_{s={\rm R,L}}\frac{k^2m^2}{4\pi M^2_{\rm P}}
\Big(z |e^{i(\frac{\pi
\nu_s}{2}+\frac{\pi}{4})}H_{\nu_s}^{(1)}(z)|^2\Big).
\end{eqnarray}
On the other hand, in the limit of $z\to 0$,  one refers to the form
of Hankel function
\begin{equation}
H^{(1)}_{\nu_s}(z\to 0)\sim
\frac{i}{\pi}\frac{\Gamma(\nu_s)}{(\frac{1}{2}z)^{\nu_s}}\Big|_{z\to
0},
\end{equation}
which implies the superhorizon limit of the vector power spectrum
\begin{eqnarray}
{\cal P}_{\Psi}\Big|_{z\to 0} &=&\sum_{s={\rm
R,L}}\frac{1}{2}\Big(\frac{2aH}{\pi}\Big)^2\Big(\frac{m}{M_{\rm
P}}\Big)^2\Big(\frac{\Gamma(\nu_s)}{\Gamma(1/2)}\Big)^2\Big(\frac{k}{2aH}\Big)^{3-2\nu_s}.
\end{eqnarray}
For $\nu_s=1/2(m^2=0)$, the power spectrum of the vector
perturbation is also zero  as
\begin{eqnarray}
{\cal P}_{\Psi}\Big|_{m^2\to0,z\to 0}= 0.
\end{eqnarray}
We wish to explain why ${\cal P}_{\Psi}$ approaches zero in the
limits of $m^2\to 0$ and $z\to 0$. In the case of $m^2\to 0$,  the
vector field becomes conformally invariant as shown in (\ref{vpeq})
and thus,  it is not gravitationally produced because it does not
couple to the expanding gravitational (de Sitter)
background~\cite{Dimopoulos:2006ms}.

In the case of $\kappa m^2=(m/M_{\rm P})^2=1$, the Weyl-squared term
in (\ref{vpeq}) reproduces a transversely massive vector Lagrangian
in (\ref{maxwell}).  For $\theta=0$, thus,  one recovers the power
spectrum ${\cal P}_{{\cal A}^\perp}$ for a transversely massive
vector ~\cite{Dimopoulos:2006ms}
\begin{eqnarray}
{\cal P}_{\Psi}\Big|_{\theta\to0,z\to 0} &=&\Big(\frac{m}{M_{\rm
P}}\Big)^2{\cal P}_{{\cal A}^\perp},
\end{eqnarray}
where \begin{equation} {\cal P}_{{\cal
A}^\perp}=\Big(\frac{2aH}{\pi}\Big)^2\Big(\frac{\Gamma(\nu_{{\cal
A}^\perp})}{\Gamma(1/2)}\Big)^2\Big(\frac{k}{2aH}\Big)^{3-2\nu_{{\cal
A}^\perp}}
\end{equation}
with \begin{equation} \nu_{{\cal
A}^\perp}=\sqrt{\frac{1}{4}-\frac{m^2_F}{H^2}}.
\end{equation}
If one defines a physical vector $V_i={\cal A}^\perp/a$, then its
power spectrum takes the form
\begin{equation} {\cal P}_{V}=\Big(\frac{2H}{\pi}\Big)^2\Big(\frac{\Gamma(\nu_{{\cal
A}^\perp})}{\Gamma(3/2)}\Big)^2\Big(\frac{k}{2aH}\Big)^{3-2\nu_{{\cal
A}^\perp}}
\end{equation}
which still vanishes in the limit of $z\to 0(k\ll aH)$ and for
$\nu_{{\cal A}^\perp}<1/2$.  Its spectral index takes the form
\begin{equation}
n_{V}=\frac{d\ln{\cal P}_{V}}{d\ln k}\simeq 2+ \frac{2m^2_F}{H^2}.
\end{equation}
for $m^2_F \ll H^2$.  Even for $m^2_F\ll H^2$, the spectrum is not
scale-invariant.

\subsection{Tensor power spectrum}

In order to derive power spectrum for tensor perturbations in the
ECSW gravity, we first rewrite the fourth-order bilinear action
(\ref{hpeq})  by using the Ostrogradsky formalism as
\begin{eqnarray}
&&\hspace*{-2.3em}\kappa S_{\rm ECSW}^{({\rm TO})}\equiv\int d^4x
{\cal L}_{\rm ECSW}^{\rm O}
\nonumber\\
&&=\frac{1}{8}\int
d^4x\Big[a^2(\alpha_{ij}\alpha^{ij}-\partial_kh_{ij}\partial^kh^{ij})
-\theta^{\prime}\epsilon_{i}^{~jk}(\alpha^{\ell
i}\partial_j\alpha_{k\ell}
-\partial^{\ell}h^{pi}\partial_j\partial_{\ell}h_{kp})\nonumber\\
&&\hspace*{7em}-\frac{1}{m^2}(\alpha_{ij}^{\prime}\alpha'^{ij}
-2\partial_k\alpha_{ij}\partial^{k}\alpha^{ij}
+\partial^2h_{ij}\partial^2h^{ij})+2\beta^{ij}(\alpha_{ij}-h'_{ij})
\Big]\label{hpeq1},
\end{eqnarray}
where $\alpha_{ij}$ is a new variable and $\beta^{ij}$ is a Lagrange
multiplier. The action (\ref{hpeq1}) is surely the second-order
bilinear action which shows that the Ostrogradsky formalism is a
well-known method to handle a higher-order action ~\cite{de
Urries:1998bi,Mannheim:2004qz,Kim:2013mfa}. Especially, the
quantization and  mean square expectation value of the nondegenerate
Pais-Uhlenbeck oscillator will be  implemented  to obtain the tensor
power spectrum~\cite{Mannheim:2004qz}.  We have two tensors $h_{ij}$
and $\alpha_{ij}$ that  amount to 4(=2+2) DOF which explains why the
fourth-order action (\ref{hpeq}) describes 4 DOF. From ${\cal
L}_{\rm ECSW}^{\rm O}$ in (\ref{hpeq1}), the conjugate momenta are
given by
\begin{eqnarray} \label{conjugate1}
\pi_h^{ij}=-\frac{1}{4\kappa}\beta^{ij},
~~~\pi_{\alpha}^{ij}=-\frac{1}{4\kappa m^2}\alpha'^{ij}.
\end{eqnarray}
The equation of motion could be obtained by varying the action
(\ref{hpeq1}) with respect to $\beta^{ij}$ and $\alpha_{ij}$ as
\begin{equation} \label{abmont}
\alpha_{ij}=h'_{ij},~~\beta_{ij}=-a^2\alpha_{ij}
-\theta^{\prime}\epsilon_{k}^{~\ell
i}\partial_{\ell}\alpha^{jk}-\frac{1}{m^2}(\alpha''_{ij}-2\partial^2\alpha_{ij}).
\end{equation}

Plugging (\ref{abmont}) into (\ref{conjugate1}), the conjugate
momenta are given by
\begin{eqnarray} \label{conjugate2}
\pi_h^{ij}=\frac{1}{4\kappa}\Big(a^2h'^{ij}
-\frac{2}{m^2}\partial^2h'^{ij} +\frac{1}{m^2}h'''^{ij}
+\theta^{\prime}\epsilon_{k}^{~\ell i}\partial_{\ell}h'^{jk}\Big),
~~~\pi_{h'}^{ij}=-\frac{1}{4\kappa m^2}h''^{ij},
\end{eqnarray}
which are the same forms obtained from the variation of (\ref{hpeq})
with respect to $h'^{ij}$ and $h''^{ij}$. Then, the corresponding
Hamiltonian is given by \begin{equation} \label {hamil} {\cal
H}_{\rm ECSW}= \pi_h^{ij} h'_{ij}+\pi_{h'}^{ij}
h''_{ij}-\tilde{{\cal L}}_{\rm ECSW}
\end{equation}
with $S_{\rm ECSW}^{(T)}=\int d^4x \tilde{{\cal L}}_{\rm ECSW}$ in
(\ref{hpeq}).

 At this stage, we would like to mention how to take
the limit of $m^2 \to 0$ in (\ref{hpeq1}). In this case, the  above
relations (\ref{conjugate1}) and (\ref{abmont}) get lost and
$h_{ij}$ becomes non-dynamical. However, taking  a $m^2\to 0$ limit
of the original tensor action (\ref{hpeq}) leads to the purely
Weyl-squared term. In the auxiliary Ostrogradsky formalism, we could
not take the $m^2\to 0$ limit directly. As was emphasized  in
Ref.~\cite{Mannheim:2004qz}, there is a crucial difference between
the general Hamiltonian  ${\cal H}^{\rm O}_{\rm ECSW}=\pi_h^{ij}
h'_{ij}+\pi_{\alpha}^{ij} \alpha'_{ij}-{\cal L}^{\rm O}_{\rm ECSW}$
and ${\cal H}_{\rm ECSW}$(\ref{hamil}). The latter Hamiltonian was
obtained when one imposes  equation (\ref{abmont}) and uses
(\ref{conjugate2}),  and it is thus called the classical and
stationary Ostrogradsky Hamiltonian which corresponds to the
fourth-order bilinear action (\ref{hpeq}). Hence, taking the limit
of $m^2\to 0$ in (\ref{hpeq1}) must be done after imposing
(\ref{abmont}) to derive the  Weyl-squared term whose gravitational
waves are not gravitationally amplified because it decoupled from
the de Sitter background. This explains why one takes the $m^2\to 0$
limit in (\ref{hpeq}) to derive a conformally invariant Weyl-squared
term.

 Now, the canonical quantization is accomplished by imposing
equal-time commutation relations:
\begin{eqnarray}\label{comm}
[\hat{h}_{ij}(\eta,{\bf x}),\hat{\pi}_{h}^{ij}(\eta,{\bf
x}^{\prime})]=2i\delta({\bf x}-{\bf
x}^{\prime}),~~~[\hat{h'}_{ij}(\eta,{\bf
x}),\hat{\pi}_{h'}^{ij}(\eta,{\bf x}^{\prime})]=2i\delta({\bf
x}-{\bf x}^{\prime}),
\end{eqnarray}
where the factor 2 is coming  from the fact that $h_{ij}$ and
$h'_{ij}(=\alpha_{ij})$ represent 2 DOF, respectively.

Here we will employ $h_{ij}$ only to compute the tensor power
spectrum by introducing two-types of mode solutions $h_{\bf
k}^{s(1)}$ and $h_{\bf k}^{s(2)}$ which correspond to Einstein and
Chern-Simons-Weyl tensor modes. Its  operator $\hat{h}_{ij}$ can be
expanded in Fourier modes as
\begin{eqnarray}\label{hex}
\hat{h}_{ij}(\eta,{\bf x})=\frac{1}{(2\pi)^{\frac{3}{2}}}\int
d^3k\Bigg[\sum_{s={\rm R,L}}\Big(\tilde{p}_{ij}^{s}({\bf
k})\hat{a}_{\bf k}^{s}h_{\bf k}^{s(1)}(\eta)e^{i{\bf k}\cdot{\bf
x}}+\tilde{p}_{ij}^{s}({\bf k})\hat{b}_{\bf k}^{s}h_{\bf
k}^{s(2)}(\eta)e^{i{\bf k}\cdot{\bf x}}\Big)+h.c.\Bigg].
\end{eqnarray}

Also, we find from (\ref{hex}) that the operator
$\hat{\pi}_{h}^{ij}$ is given by
\begin{eqnarray}\label{piex}
&&\hspace*{-3em}\hat{\pi}_{ij}^h(\eta,{\bf
x})=\frac{1}{(2\pi)^{\frac{3}{2}}}\int
d^3k\frac{1}{4\kappa}\Bigg[\sum_{s={\rm
R,L}}\Bigg(\tilde{p}_{ij}^{s}({\bf k})\hat{a}_{\bf
k}^{s}\Big\{\xi^s\Big(h_{\bf
k}^{s(1)}(\eta)\Big)^{\prime}+\frac{1}{m^2}\Big(h_{\bf
k}^{s(1)}(\eta)\Big)^{\prime\prime\prime}\Big\}e^{i{\bf k}\cdot{\bf
x}}\nonumber\\
&&\hspace*{4.4em}+~\tilde{p}_{ij}^{s}({\bf k})\hat{b}_{\bf
k}^{s}\Big\{\xi^s\Big(h_{\bf
k}^{s(2)}(\eta)\Big)^{\prime}+\frac{1}{m^2}\Big(h_{\bf
k}^{s(2)}(\eta)\Big)^{\prime\prime\prime}\Big\}e^{i{\bf k}\cdot{\bf
x}}\Bigg)+h.c.\Bigg],
\end{eqnarray}
where $\xi^s$ is
\begin{eqnarray}
\xi^s=a^2+\frac{2}{m^2}k^2-\lambda^sk\theta^{\prime}.
\end{eqnarray}
On the other hand, $h'_{ij}$ is  given by
\begin{eqnarray}\label{hex2}
\hat{h}'_{ij}(\eta,{\bf x})&=&\frac{1}{(2\pi)^{\frac{3}{2}}}\int
d^3k\Bigg[\sum_{s={\rm R,L}}\Big(\tilde{p}_{ij}^{s}({\bf
k})\hat{a}_{\bf k}^{s}\Big(h_{\bf k}^{s(1)}(\eta)\Big)'e^{i{\bf
k}\cdot{\bf x}} \nonumber \\
&+&\tilde{p}_{ij}^{s}({\bf k})\hat{b}_{\bf k}^{s}\Big(h_{\bf
k}^{s(2)}(\eta)\Big)'e^{i{\bf k}\cdot{\bf x}}\Big)+h.c.\Bigg],
\end{eqnarray}
and  $\pi_{h'ij}$ takes the form
\begin{eqnarray}\label{hpex2}
\hat{\pi}_{h'ij}(\eta,{\bf
x})&=&-\frac{1}{(2\pi)^{\frac{3}{2}}}\int d^3k\frac{1}{4\kappa
m^2}\Bigg[\sum_{s={\rm R,L}}\Big(\tilde{p}_{ij}^{s}({\bf
k})\hat{a}_{\bf k}^{s}\Big(h_{\bf k}^{s(1)}(\eta)\Big)''e^{i{\bf
k}\cdot{\bf x}}\nonumber \\
&+&\tilde{p}_{ij}^{s}({\bf k})\hat{b}_{\bf k}^{s}\Big(h_{\bf
k}^{s(2)}(\eta)\Big)''e^{i{\bf k}\cdot{\bf x}}\Big)+h.c.\Bigg].
\end{eqnarray}
Substituting (\ref{hex}), (\ref{piex}), (\ref{hex2}), and
(\ref{hpex2}) into (\ref{comm}) leads to the consistent commutation
relations and Wronskian conditions:
\begin{eqnarray}
&&\hspace*{-2em}[\hat{a}_{\bf k}^{s},\hat{a}_{\bf k^{\prime}}^{
s^{\prime}\dag}]=\delta^{ss^{\prime}}\delta({\bf k}-{\bf
k}^{\prime}),~~~~~~[\hat{b}_{\bf k}^{s},\hat{b}_{\bf
k^{\prime}}^{s^{\prime}\dag}]=-\delta^{ss^{\prime}}\delta({\bf
k}-{\bf
k}^{\prime}),\label{com-rel2}\\
\nonumber\\
&&\hspace*{-2em}\Big[h_{\bf k}^{s(1)}\Big\{\xi^s(h_{\bf
k}^{*s(1)})^{\prime}+\frac{1}{m^2}(h_{\bf
k}^{*s(1)})^{\prime\prime\prime}\Big\}-h_{\bf
k}^{s(2)}\Big\{\xi^s(h_{\bf
k}^{*s(2)})^{\prime}+\frac{1}{m^2}(h_{\bf
k}^{*s(2)})^{\prime\prime\prime}\Big\}\Big]-c.c.=4i\kappa ,\label{wcon}\\
\nonumber\\
&&\hspace*{-2em}\Big[(h_{\bf k}^{s(1)})^{\prime}(h_{\bf
k}^{*s(1)})^{\prime\prime}-(h_{\bf k}^{s(2)})^{\prime}h_{\bf
k}^{*s(2)})^{\prime\prime}\Big]-c.c.=-4i\kappa m^2,\label{wcon1}
\end{eqnarray}
where $\tilde{p}_{ij}^{\rm L}(\tilde{p}^{ij\rm
L})^*=\tilde{p}_{ij}^{\rm R}(\tilde{p}^{ij\rm R})^*=1$ and a
superscript ``$s$'' in (\ref{wcon}) and (\ref{wcon1}) does not
denote summation over $({\rm L,R})$. We note that two mode operators
($\hat{a}_{\bf k}^{s},\hat{b}_{\bf k}^{s}$) are needed to take into
account the fourth-order bilinear tensor action (\ref{hpeq1}) by
using the Ostrogradsky formalism.

 Importantly, introducing
$\mu_{\bf k}^{s({\rm E})}=\mu_{\bf k}^{s(1)}$ and $\mu_{\bf
k}^{s({\rm CSW})}=\mu_{\bf k}^{s(2)}$, Eqs.(\ref{hee}) and
(\ref{hcsw}) in the de Sitter background with $z=-\eta k$ can be
written by (\ref{hc0}) and (\ref{hc11}), respectively.
 The consistency condition to satisfy
Eqs. (\ref{hc0})-(\ref{hc11}) is given by
\begin{eqnarray}
h_{\bf k}^{s(1)}\frac{dh_{\bf k}^{*s(1)}}{dz}-c.c.=h_{\bf
k}^{s(2)}\frac{dh_{\bf k}^{*s(2)}}{dz}-c.c.=-\frac{4i\kappa m^2
z^2}{k^3(2+\tilde{m}_{s}^{2}/H^2)},
\end{eqnarray}
where $\tilde{m}_{s}^{2}$ is given by $\tilde{m}_{R/L}^{2}=m^2(1\pm
kH^2M_{\rm P}^{-3})$. We note that if two modes have the same
normalization, we could not determine their normalization constants
because Eqs.(\ref{wcon}) and (\ref{wcon1}) provides the same
relation.

It turns out that when we consider the initial condition to set the
Bunch-Davies vacuum, the two solutions are given by
\begin{eqnarray}
h_{\bf k}^{s(1)}&=&\sqrt{\frac{2 \kappa
m^2}{k^3(2+\tilde{m}_s^2/H^2)}}\sqrt{\frac{\pi}{2}}e^{i\pi}z^{\frac{3}{2}}H_{3/2}^{(1)}(z),
\label{s1f}\\
h_{\bf k}^{s(2)}&=&\sqrt{\frac{2\kappa
m^2}{k^3(2+\tilde{m}_s^2/H^2)}} \sqrt{\frac{\pi}{2}}e^{i(\frac{\pi
\nu_s}{2}+\frac{\pi}{4})}z^{\frac{3}{2}}H_{\nu_s}^{(1)}(z),\label{s2f}
\end{eqnarray}
where $H_{\nu_s}^{(1)}$ is the Hankel function and here $\nu_s$ is
given by (\ref{ns}). Comparing these solutions with
(\ref{n-tensor}), the normalizations  are fixed in the former cases.

 On the other hand, the power spectrum of the
gravitational waves is defined by
\begin{eqnarray}\label{power}
\langle0|\hat{h}_{ij}(\eta,{\bf x})\hat{h}^{ij}(\eta,{\bf
x^{\prime}})|0\rangle=\int d^3{\bf k}\frac{{\cal P}_{\rm T}}{4\pi
k^3}e^{i{\bf k}\cdot({\bf x}-{\bf x^{\prime}})},
\end{eqnarray}
where we choose  the Bunch-Davies vacuum state $|0>$ by imposing
$\hat{a}_{\bf k}^{s}|0\rangle=0$ and $\hat{b}_{\bf
k}^{s}|0\rangle=0$.  A quantity ${\cal P}_{\rm T}$ in
(\ref{power}) denotes ${\cal P}_{\rm T}\equiv\sum_{s={\rm
R,L}}{\cal P}^s$ given as
\begin{eqnarray} \label{ten-power}
{\cal P}^s=\frac{k^3}{2\pi^2}\left(\Big|h_{\bf
k}^{s(1)}\Big|^2-\Big|h_{\bf k}^{s(2)}\Big|^2\right),
\end{eqnarray}
where the difference ($-$) arises from the commutation relations
(\ref{com-rel2}).

It is very important to note that the tensor power spectrum
(\ref{ten-power}) is based on the factorization (\ref{dec1}) and
(\ref{dec2}) which can be realized only when $\theta$ takes the form
$\theta=c/\eta$. Unless choosing this $\theta$, it is difficult to
obtain the corresponding tensor power spectrum. Substituting
(\ref{s1f}) and (\ref{s2f}) into (\ref{power}), we obtain with
$\kappa=1/M^2_{\rm P}$
\begin{eqnarray} \label{ten-power1}
{\cal P}_{\rm T}=\sum_{s={\rm R,L}}\frac{ H^2}{\pi^2M^2_{\rm
P}}\frac{m^2}{\tilde{m}_s^2+2H^2} \Big(1+z^2-\frac{\pi}{2}z^3 |e^{i
(\frac{\pi \nu_s}{2}+\frac{\pi}{4})} H_{\nu_s}^{(1)}(z)|^2\Big).
\end{eqnarray}
In the limit of $m^2\to 0$ (keeping Weyl-squared term only), one has
$\nu_s \to 1/2$ and $\tilde{m}^2 \to 0$. This case provides the zero
power spectrum as \begin{eqnarray} \label{ten-powerm} {\cal P}_{\rm
T}\Big|_{m^2\to 0} \to 0
\end{eqnarray}
which implies that the tensor perturbation becomes conformally
invariant and thus, they are not gravitationally produced because
they are decoupled from the expanding de Sitter background as was
shown in (\ref{hpeq}).

 In the case of Einstein-Weyl gravity with
$\theta=0(\tilde{m}_s^2=m^2)$, the tensor power spectrum
(\ref{ten-power1}) reduces to
\begin{equation} \label{ten-power2}
{\cal P}^{\rm EW}_{\rm T}={\cal P}^{\rm GW}\frac{m^2}{m^2+2H^2}
\Big(1+z^2-\frac{\pi}{2}z^3
|e^{i(\frac{\pi\nu}{2}+\frac{\pi}{4})}H_{\nu}^{(1)}(z)|^2\Big),
\end{equation}
where ${\cal P}^{\rm GW}=\frac{2 H^2}{\pi^2M^2_{\rm P}}$ is the
power spectrum for gravitational waves and $\nu=\sqrt{1/4-m^2/H^2}$.
This is  the same power spectrum obtained in
Ref.\cite{Deruelle:2012xv}.

In the superhorizon limit of $z\to 0$, the second in
(\ref{ten-power1}) is zero and the third term approaches zero as
\begin{eqnarray}
\frac{\pi}{2}z^3 |e^{i(\frac{\pi \nu_s}{2}+\frac{\pi}{4})}
H_{\nu_s}^{(1)}|^2\Big|_{z\to
0}=\frac{1}{2\pi}\Gamma(\nu_s)^22^{2\nu_s}z^{3-2\nu_s}\Big|_{z\to 0}
\to 0
\end{eqnarray}
for $\nu_s<1/2$ which is confirmed from (\ref{ns}). In this case,
(\ref{ten-power1})
 leads to the power spectrum for GWs
\begin{eqnarray}
{\cal P}_{\rm T}\Big|_{z\to 0}&=&\sum_{s={\rm
R,L}}\frac{H^2}{\pi^2M^2_{\rm
P}}\Big[\frac{m^2}{\tilde{m}_s^2+2H^2}\Big]\nonumber\\
&=& {\cal P}^{\rm GW}\frac{(1+2H^2/m^2)}{(1+2H^2/m^2)^2-(kH^2/M_{\rm
P}^3)^2} \equiv {\cal P}^{\rm GW}\Xi_{\rm CSW},
\end{eqnarray}
where one recovers $\Xi_{\rm EW}=\frac{1}{1+2H^2/m^2}$ in ${\cal
P}^{\rm EW}_{\rm T}|_{z\to 0}={\cal P}^{\rm GW}\Xi_{\rm EW}$ of the
Einstein-Weyl gravity~\cite{Deruelle:2012xv}. We note that ${\cal
P}_{\rm T}|_{z\to 0}$ is not a scale-invariant spectrum, but ${\cal
P}^{\rm EW}_{\rm T}|_{z\to 0}$ is a scale-invariant spectrum.
Considering the bound (\ref{mass-b}), we have $2H^2/m^2>8$. Also, we
have $kH^2/M^3_{\rm P} \ll 1$. In this case, the mass-squared term
damps out smoothly the spectrum of primordial gravitational waves
$({\cal P}^{\rm GW})$ because
\begin{equation}
 {\cal P}_{\rm
T}\Big|_{z\to 0}<\frac{1}{9}{\cal P}^{\rm GW}.
\end{equation}

Finally, we would like to mention  the massive tensor equation from
the general massive gravity~\cite{Bessada:2009np}
\begin{equation}
h_{ij}''+2{\cal H}h_{ij}'+m^2_2a^2h_{ij}-\partial^2h_{ij}=0,
\end{equation}
which is the same equation (\ref{inflaton-eq1}) as for the massive
scalar $\delta \phi$. Considering the normalization of $\delta \phi
\to \frac{M_{\rm P}}{2}h_{ij}$, the power spectrum is given by
\begin{equation} \label{gmgml0}
{\cal P}_{\rm GMG}=2\times \Big(\frac{2}{M_{\rm P}}\Big)^2 {\cal
P}_{\delta\phi}\Big|_{\nu_\phi \to \nu_{m_2}}=\frac{H^2}{\pi
M^2_{\rm P}}z^3|e^{i(\frac{\pi
\nu_{m_2}}{2}+\frac{\pi}{4})}H^{(1)}_{\nu_{m_2}}(z)|^2,
\nu_{m_2}=\sqrt{\frac{9}{4}-\frac{m^2_2}{H^2}}
\end{equation}
with ${\cal P}_{\delta\phi}$ (\ref{msp0}). In the limit of $z\to 0$
and $m^2_2/H^2 \ll 1$, one has the massive tensor power spectrum
\begin{equation}
{\cal P}_{\rm GMG}\simeq \Big(\frac{2H^2}{\pi^2 M^2_{\rm
P}}\Big)\Big(\frac{k}{2aH}\Big)^{\frac{2m^2_2}{3H^2}}
\end{equation}
whose spectral index takes the form
\begin{equation}
n_{\rm GMG}=\frac{d\ln{\cal P}_{\rm GMG}}{d\ln k}\simeq
\frac{2m^2_2}{3H^2}.
\end{equation}
In the case of $m^2_2=0$, we have the power spectrum for GWs as and
its spectral index
\begin{equation} \label{gmgml1}
{\cal P}_{\rm GMG}\Big|_{m^2_2 \to 0} = \Big(\frac{2H^2}{\pi^2
M^2_{\rm P}}\Big),~~n_{\rm GMG}\Big|_{m^2_2 \to 0}=0.
\end{equation}

\section{Discussions}

In the  Einstein gravity, all fluctuations of scalar and tensor
originate on subhorizon scales and they propagate for a long time on
superhorizon scales. There is no vector propagation.  This can be
checked by computing their power spectra in the superhorizon limit
of $z\to 0$. However, we have found the different results for
massive fluctuations composed of scalar $\Phi=\Psi$,  vector
$\Psi_i$ with 2 DOF, and tensor $h_{ij}$ with 4 DOF.

First of all, we have  derived a power spectrum ${\cal P}_\Phi$ for
the scalar $\Phi$ which blows up  in the superhorizon limit. In the
case of Einstein gravity, however, $\Phi$ was combined to give
comoving curvature perturbation ${\cal R}$ on the slow-roll
inflation. Also, we have obtained a scale-variant power spectrum
${\cal P}_{\delta \phi}$ for a massive scalar $\delta \phi$.

The power spectra of massive vector $\Psi_i$ shows clearly that
their fluctuations do not propagate for a long time on superhorizon
scales. It decays quickly  in the superhorizon limit of $z\to 0$.
Also, in the limit of $m^2\to 0$ (keeping the Weyl-squared term
only), it disappears because  the vector field becomes conformally
invariant as shown in (\ref{vpeq}). It is not gravitationally
produced because it does not couple to the expanding gravitational
(de Sitter) background~\cite{Dimopoulos:2006ms}. There exist close
connection in power spectra between $\{ \Psi_i, \delta \phi\}$ and
$\{ {\cal A}^\perp,{\cal A}^\parallel\}$, where $\delta \phi$ a
massive scalar, ${\cal A}^\perp$ and ${\cal A}^\parallel$ are
 a transversely massive vector and longitudinally light  massive vector,
 respectively~\cite{Dimopoulos:2006ms}. We have found that
 ${\cal P}_\Psi \simeq {\cal P}_{{\cal A}^\perp}$ and ${\cal P}_{\delta \phi} \simeq {\cal P}_{{\cal
 A}^\parallel}$.  In order to compute ${\cal P}_\Psi$, however,  we have chosen
 the conventional conjugate momentum $\pi_\Psi=\frac{1}{2\kappa m^2}(\Psi^j)'$
 instead of $\tilde{\pi}_\Psi=\frac{1}{2\kappa m^2}\partial^2 (\Psi^j)'$ obtained
 from (\ref{vpeq}). The latter induces an inconsistent quantization
 of $[\hat{a}^s_{\bf k},\hat{a}^{s'}_{{\bf
 k}'}]=-\delta^{ss'}\delta({\bf k}-{\bf k}')$ and thus, an unconventional power spectrum.

The power spectrum of tensor has taken a complicated from
(\ref{ten-power1}) because it was obtained by using the Ostrogradsky
formalism to handle the fourth-order theory. In this case, one
usually introduces two sets of lowering and raising operator for all
${\bf k}$. Here, the ghost state problem might appear because the
tensor equation is a fourth-order equation with respect to conformal
time $\eta$. In the superhorizon limit of $z\to 0$, however, the
tensor power spectrum coming from massive GWs  (2 DOF) decays
quickly, remaining the conventional tensor (2 DOF) power spectrum.
Importantly, in the limit of $m^2\to 0$ (keeping the Weyl-squared
term only), the tensor power spectrum (\ref{ten-power1}) disappears
because the tensor  field becomes conformally invariant as shown in
(\ref{hpeq}). It is not gravitationally produced because it is
decoupled from  the expanding gravitational (de Sitter) background.
However, considering the power spectrum (\ref{gmgml0}) from the
general massive gravity, its massless limit (\ref{gmgml1}) recovers
the conventional tensor power spectrum and spectral index. This
implies that a different massive gravity provides a different power
spectrum.

Finally,  we would like to mention the  choice of Chern-Simons
scalar $\theta$. It remains undetermined in the de Sitter inflation.
For $\theta=c\ln \eta$,  Eq.(\ref{slowms-eq}) could describe
propagation of circularly polarized GWs because the Chern-Simons
term is regarded  as a small correction.  However, their power
spectrum is the same as that of GWs on the de Sitter
inflation~\cite{Satoh:2008ck}. For $\theta=c/\eta$, however, one
make  factorization of fourth-order tensor equation which is a
necessary step to compute the tensor power spectrum. In this work,
we have chosen  $\theta=c/\eta$ to obtain two second-order tensor
equations (\ref{hc0}) and (\ref{hc11}) because $\theta=c\ln \eta$
unlikely factorize the fourth-order equation into two second-order
equations.

\vspace{0.25cm}

 {\bf Acknowledgement}

\vspace{0.25cm}
 This work was supported by the National
Research Foundation of Korea (NRF) grant funded by the Korea
government (MEST) (No.2012-R1A1A2A10040499).

\newpage

\section*{Appendix: Factorization of the fourth-order tensor equation}

In the de Sitter background with $z=-\eta k$, the fourth-order
differential equation (\ref{heq2}) can be rewritten as
\begin{eqnarray}
&&\hspace*{-2em}\frac{d^4}{dz^4}h_{\bf
k}^{s}(z)+2\frac{d^2}{dz^2}h_{\bf k}^{s}(z)+h_{\bf k}^{s}(z)
+m^2\Bigg\{\Big(\frac{1}{H^2z^2}+\lambda^s\frac{d\theta}{dz}\Big)
\frac{d^2}{dz^2}h_{\bf
k}^{s}(z)\nonumber\\
&&\hspace*{4.5em}-\Big(\frac{2}{H^2z^3}-\lambda^s\frac{d^2\theta}{dz^2}\Big)
\frac{d}{dz}h_{\bf
k}^{s}(z)+\Big(\frac{1}{H^2z^2}+\lambda^s\frac{d\theta}{dz}\Big)
h_{\bf k}^{s}(z)\Bigg\}=0.\label{foth}
\end{eqnarray}
To factorize the above fourth-order equation (\ref{foth}) into two
second-order equations, we consider a factorized form
\begin{eqnarray}
\Big(X_1(z)\frac{d^2}{dz^2}+X_2(z)\frac{d}{dz}+X_3(z)\Big)
\Big(X_4(z)\frac{d^2}{dz^2}+X_5(z)\frac{d}{dz}+X_6(z)\Big)h_{\bf
k}^{s}(z)=0.\label{sech}
\end{eqnarray}
Comparing the coefficients of $\frac{d^2}{dz^2}h_{\bf
k}^{s}(z),~\frac{d^3}{dz^3}h_{\bf k}^{s}(z),~\frac{d^4}{dz^4}h_{\bf
k}^{s}(z)$ in (\ref{foth}) and (\ref{sech}), we find
\begin{eqnarray}
&&\hspace*{-2em}X_1=\frac{1}{X_4},~~
X_2=\frac{2}{zX_4}-\frac{2}{X_4^2}\frac{dX_4}{dz},\label{x1}\\
\nonumber\\
&&\hspace*{-2em}X_3=\frac{1}{X_4}\Bigg\{2-\frac{X_6}{X_4}-\frac{2}{zX_4}\frac{dX_4}{dz}
+\frac{2}{X_4^2}\Big(\frac{dX_4}{dz}\Big)^2-\frac{1}{X_4}\frac{d^2X_4}{dz^2}
+m^2\Big(\frac{1}{H^2z^2
}+\lambda^s\frac{d\theta}{dz}\Big)\Bigg\},\label{x2}
\end{eqnarray}
where we used a condition of $X_5=-2X_4/z$, which yields the
Schr\"odinger-type equation for $\mu_{\bf k}^{s}$($=h_{\bf
k}^{s}/z$). Substituting Eqs. (\ref{x1}) and (\ref{x2}) into
(\ref{sech}) and comparing the coefficients of $h_{\bf k}^{s}(z)$
and $\frac{d}{dz}h_{\bf k}^{s}(z)$ in (\ref{foth}) and (\ref{sech})
leads to
\begin{eqnarray}
&&\hspace*{-5em}1+\frac{m^2}{H^2z^2}+m^2\lambda^s\frac{d\theta}{dz}
-\Big(2+\frac{m^2}{H^2z^2}+m^2\lambda^s\frac{d\theta}{dz}\Big)
\frac{X_6}{X_4}+\frac{X_6^2}{X_4^2}\nonumber\\
&&\hspace*{12em}-\frac{2}{z}\frac{d}{dz}\Big(\frac{X_6}{X_4}\Big)
-\frac{d^2}{dz^2}\Big(\frac{X_6}{X_4}\Big)=0,\label{cou1}\\
&&\nonumber\\
&&\frac{4}{z}-\frac{4X_6}{zX_4}-2\frac{d}{dz}\Big(\frac{X_6}{X_4}\Big)
+m^2\lambda^s\Big(\frac{2}{z}\frac{d\theta}{dz}+\frac{d^2\theta}{dz^2}\Big)=0.\label{cou2}
\end{eqnarray}
It turns out that two coupled  equations (\ref{cou1}) and
(\ref{cou2}) for $X_6/X_4$ and $\theta$ can be solved by choosing
two cases:
\begin{eqnarray}
{\rm case~1}:&&\frac{X_6}{X_4}=1,~~~\theta=c_2-\frac{c_1}{z}\\
&&\nonumber\\
{\rm case~2}:&&\frac{X_6}{X_4}=1+\Big(2+c_3
m^2\lambda^s+\frac{m^2}{H^2}\Big)\frac{1}{z^2},~~~\theta=c_4-\frac{c_3}{z},
\end{eqnarray}
where $c_i$ with $i=1,2,3,4$ are integration constants.
Choosing\footnote{We note that in the Einstein-Weyl gravity limit
($\lambda^s\to0$), these choices of $X_4$ give the consistent
factorization obtained in the Einstein-Weyl gravity
~\cite{Deruelle:2010kf}} $X_4=1$ in case 1 and $X_4=z^2$ in case 2,
the variables $X_i$ with $i=1,\cdots,6$ in Eq. (\ref{sech}) are
determined to provide
\begin{eqnarray} \label{twoeq1}
\Bigg[\frac{d^2}{dz^2}+\frac{2}{z}\frac{d}{dz}+1+m^2\Big(\frac{1}{z^2H^2}
+\lambda^s\frac{d\theta}{dz}\Big)\Bigg]\Bigg[
\frac{d^2}{dz^2}-\frac{2}{z}\frac{d}{dz}+1\Bigg]h_{\bf k}^{s}(z)=0
\end{eqnarray}
and
\begin{eqnarray} \label{twoeq2}
\Bigg[\frac{1}{z^2}\frac{d^2}{dz^2}-\frac{2}{z^3}\frac{d}{dz}+\frac{1}{z^2}\Bigg]\Bigg[
z^2\frac{d^2}{dz^2}-2z\frac{d}{dz}+2+z^2+m^2\Big(\frac{1}{H^2}
+\lambda^sz^2\frac{d\theta}{dz}\Big)\Bigg]h_{\bf k}^{s}(z)=0,
\end{eqnarray}
which are Eqs. (\ref{dec1}) and (\ref{dec2}), respectively. This
implies that the choice of $\theta=c_2-c_1/z$ could transform the
fourth-order equation (\ref{foth}) into the product of two
second-order equations as (\ref{twoeq1}) and (\ref{twoeq2}). This
factorization unlikely occurs for $\theta=c \ln z$.   The same thing
happens when one uses $\eta$ instead of $z$. This is why we choose
$\theta=c/\eta$ for the analysis in the text.

\end{document}